\documentclass[twocolumn,showpacs,preprintnumbers,amsmath,amssymb,final]{revtex4}

\usepackage{graphicx}% Include figure files
\usepackage{dcolumn}% Align table columns on decimal point
\usepackage{bm}% bold math

\def\BEq{\begin{equation}}
\def\EEq{\end{equation}}
\def\BEqA{\begin{eqnarray}}
\def\EEqA{\end{eqnarray}}
\def\BEn{\begin{enumerate}}
\def\EEn{\end{enumerate}}
\def\BWT{\begin{widetext}}
\def\EWT{\end{widetext}}
\begin{document}

%\preprint{APS/123-QED}

\title{ High-energy cosmic ray production
by a neutron star falling into a black hole}

%\author{Andrei Galiautdinov}
% \affiliation{Department of Physics and Astronomy, 
%University of Georgia, Athens, Georgia 30602, USA}

\author{Andrei Galiautdinov$^1$ and David Finkelstein$^2$}
 \affiliation{$^1$Department of Physics and Astronomy, 
University of Georgia, Athens, GA 30602, USA \\
$^2$School of Physics, 
Georgia Institute of Technology, Atlanta, GA 30332, USA}

\date{\today}% It is always \today, today,

\begin{abstract}

{{
We propose a
one-shot
mechanism for
high-energy cosmic ray generation by a neutron star falling 
into a black hole surrounded by low density plasma.}} 
{{The function of the black hole in this scenario is to accelerate the 
star to a speed {\it arbitrarily} close to that of light. When the star
--- essentially, a magnetized sphere --- approaches the horizon it imparts 
energy to the ambient plasma charges via the induced electric field.}}
Disregarding radiation losses, for iron nucleus, a simple estimate gives 
energies on the order of $10^{19}$ eV for stars with magnetic fields 
as weak as $10^6$ teslas. The proposed mechanism should also work 
in chance encounters between rapidly moving neutron stars and molecular clouds.
The rarity of such encounters may explain 
the apparent 
randomness and rarity of the high-energy cosmic ray events.

\end{abstract}


\maketitle

%\tableofcontents

%\section{Introduction}

The origin of high- and ultra-high-energy (HE and UHE, respectively) cosmic rays --- charged nuclei with energies up to and exceeding $10^{18}$ eV arriving to Earth from outer space ---
has been the subject of numerous discussions
for more than fifty years \cite{Troitsky2013, Torres2004, NaganoWatson2000,
Berezinsky1990}. 
Several mechanisms explaining 
the acceleration process
have been put forward 
\cite{LS2011}, chief among them
being diffusive shock acceleration based on the Fermi mechanism
\cite{Fermi1949}, and 
one-shot acceleration in a strong electric field induced
by a rapidly rotating pulsar \cite{NeronovSemikoz2012}.
Here we propose another one-shot mechanism 
to accelerate charges to ultra-high energies. 

Envision a magnetized neutron star on a collision course with
a black hole surrounded by a low density plasma. 
An observer hovering above the horizon suddenly sees the
star rapidly flying by.
The changing magnetic field $B$ at {{the}} observer's location
induces an electric field {{that}} accelerates the
particles of the surrounding plasma.
Acceleration of a given charge lasts for only a brief time,
$\Delta t \simeq R/v$, where $R$ is the radius of the neutron star
and $v$ is its speed relative to the hovering observer, which near the horizon 
approaches the speed of light, $c$. 
The work $W_q$ done by the induced electric 
field $E$ on a charge $q$ is then, roughly,
\BEq
W_q 
\simeq q E c \Delta t 
\simeq q E R \simeq q\frac{\Delta B }{\Delta t}R^2,
\EEq
where in the application of Faraday's law we have chosen an Amperian loop 
in the form of a circle of radius $R$. Taking $\Delta B \simeq B$, we get an estimate
\BEq
\label{eq:estimate}
W_q \simeq qBRc.
\EEq 
{{For}} iron, with
$q = 26\times 1.6\times 10^{-19} {\rm \,C}$, and the star with 
$R \simeq  5\times 10^3 {\rm \, m}$, even for a relatively weak magnetic field 
of $B \simeq  10^6 {\rm \,T}$, this gives particle
energies on the order of
$W_q \simeq 1 {\rm \, J} \simeq 10^{19} {\rm \, eV}$.
This estimate agrees with the Hillas criterion for one-shot 
cosmic ray production \cite{Hillas1984};
it also follows from dimensional analysis: 
Eq.\ (\ref{eq:estimate}) is the simplest reasonable combination of 
quantities playing a role in this scenario that has units of energy. 

Like any other induction-based one-shot acceleration scenario, 
our mechanism suffers from the limitations imposed by    
radiation losses
\cite{Medvedev2003,NTT2005,PtitsynaTroitsky2010}.
For the iron nucleus, the
maximum attainable energy in the synchrotron-loss-saturated regime
under the conditions quoted above is given by
\BEq
{\cal E}_{\rm syn}
= \sqrt{
\frac{3}{2}\frac{4\pi \epsilon_0(mc^2)^4}
{q^3 cB}} \simeq 10^{16} {\rm eV},
\EEq
where $\epsilon_0$ is the permittivity of free space, and $m$ is the mass
of the nucleus
\cite{Medvedev2003}. 
However, since UHE cosmic ray production is ultimately a quantum 
mechanical process, its probabilistic nature may allow the occurrence of rare
events at energies much higher than ${\cal E}_{\rm syn}$.
Additionally, there may also be a possibility of a purely kinematical resolution 
of the above mentioned limit.

\begin{figure}[!ht]
\includegraphics[angle=0,width=1.00\linewidth]{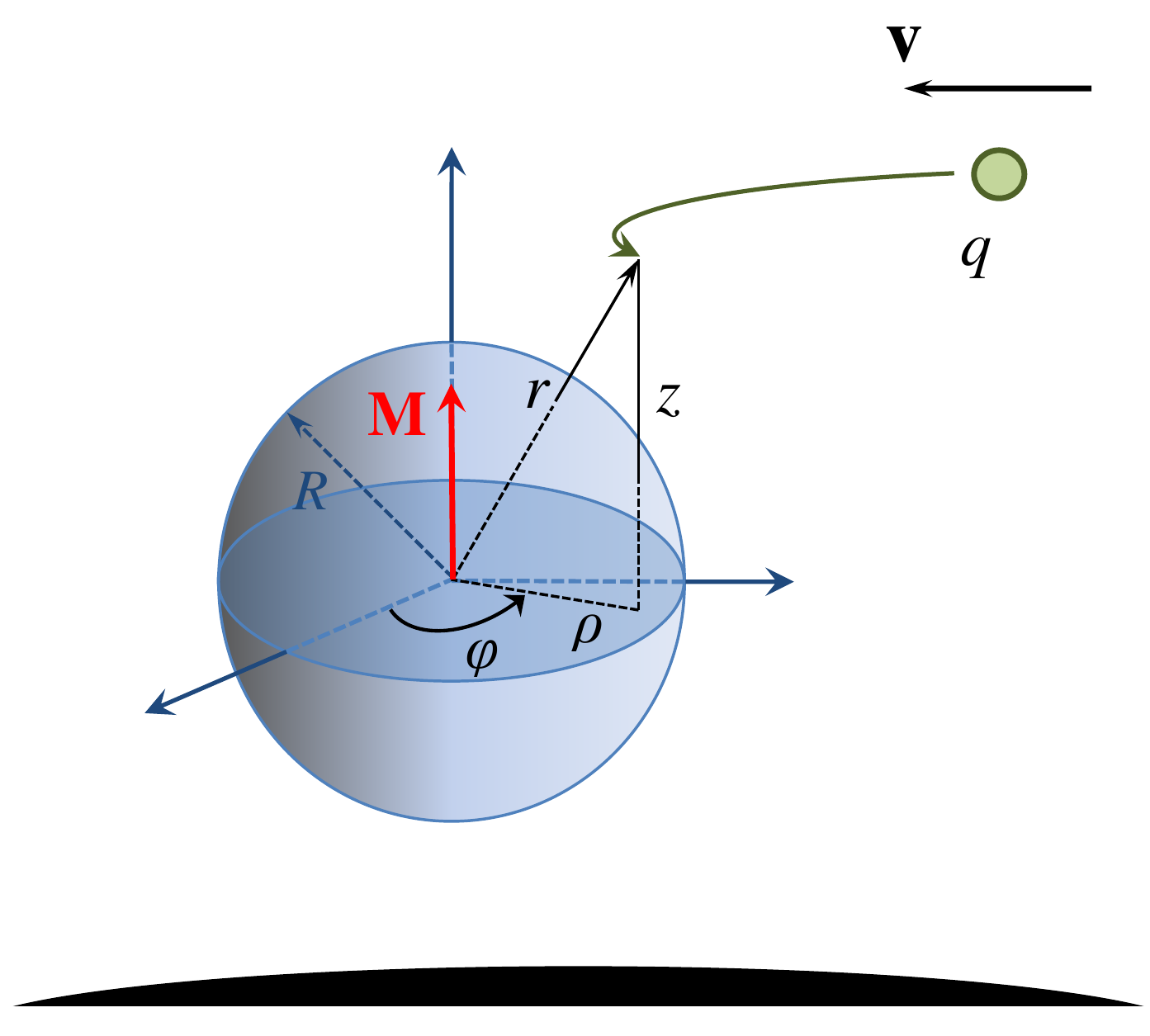}
\caption{ \label{fig:1} 
(color online). 
Schematic representation of motion of a charged particle $q$ as viewed 
from the reference frame of the neutron star (depicted here as a magnetized 
sphere of radius $R$ having
the magnetic dipole moment ${\bf M}$). 
The black hole horizon is shown in black. 
The sideways kick received by the charge during the encounter is the mechanism 
behind the ultra-high-energy cosmic ray production. Gravitational 
forces acting on the charge during this encounter are ignored.
}
\end{figure}

Thus, disregarding the radiation constraint, let us take a closer look at the
proposed acceleration mechanism by performing a
quantitative analysis of particle motion in 
the field of a rapidly approaching magnetic dipole. The process
is schematically shown in Fig.\ \ref{fig:1}. The star is modeled as a 
magnetized sphere whose dipole moment ${\bf M}$ is assumed to 
be perpendicular to the direction of its propagation. 
{{In the frame of the star,}} the charge is 
seen as impinging on the magnetic dipole 
at ultra-relativistic speed, $v\simeq c$. The sideways kick it receives 
due to the Lorentz force $q{\bf v} \times {\bf B}$ is the proposed mechanism 
{{for}} the cosmic ray production. 

%%%%%%%%% BEGIN FIG. 1
%%%%%%%%% END FIG. 1

 A curious conceptual analogy is in order here. 
The problem of {{the}} motion of charges in the field of a magnetic dipole 
has a long history
and is known as the St{\"o}rmer Problem \cite{{Alfven1963},Saletan1998}. 
It forms the basis of the theory of polar auroras in 
Earth's atmosphere \cite{Stormer1955}. 
It is used to explain the mechanism by which 
the incoming cosmic rays get trapped by the magnetic field of the Earth. Our
scenario then is the opposite of the aurora (call it ``anti-aurora'') ---
the charges, instead of being trapped, get scattered {\it away} from the dipole.

In cylindrical coordinates, the Lagrangian $L$ of a particle with mass $m$
and charge $q$ moving with speed $v$ in the 
magnetic field of the dipole ${\bf M} = M \hat{\bf z}$ is given by
\BEq
L = -m c^2 \sqrt{1-\frac{v^2}{c^2}} 
+ \frac{\mu_0}{4\pi}\frac{qM\rho^2\dot{\varphi}}{(\rho^2+z^2)^{3/2}},
\EEq  
where $\mu_0$ is the permeability of free space, and the over-dot indicates 
differentiation with respect to time $t$. There are two constants of motion in this 
problem: the speed,
\BEq
\label{eq:v}
v^2 = \dot{\rho}^2+ \rho^2\dot{\varphi}^2+\dot{z}^2,
\EEq
and the generalized momentum,
\BEq
\label{eq:pvarphi}
p_{\varphi} = \gamma m\rho^2\dot{\varphi}+ 
\frac{\mu_0}{4\pi}\frac{qM\rho^2}{(\rho^2+z^2)^{3/2}},
\quad 
\gamma \equiv [1-v^2/c^2]^{-1/2},
\EEq
which at large distance has the meaning of the usual angular momentum with
respect to the $z$ axis. In the so-called ``equatorial limit'' (corresponding to the initial 
conditions $z(0)=0$, $\dot{z}(0)=0$, which lead to $z(t)=0$ for all $t$), particle
motion in the radial direction is described by the effective Lagrangian
\BEq
L_{\rm eff} = \frac{1}{2}{\gamma} m\dot{\rho}^2 - 
\frac{1}{2{\gamma}m\rho^2}
\left(
p_{\varphi}-\frac{\mu_0}{4\pi}\frac{qM}{\rho}
\right)^2.
\EEq
From this, for a head-on collision with $p_{\varphi}=0$, we can find the 
point of closest approach,
\BEq
\label{eq:rhoMin}
\rho_{\rm min} = \sqrt{\frac{\mu_0}{4\pi}\frac{qM}{\gamma mv}}.
\EEq
If we now take into account the relation between the dipole moment $M$ of 
the sphere and the magnetic field $B_0$ on the equator,
\BEq
\label{eq:MvsB0}
M = \frac{4\pi}{\mu_0}B_0 R^3,
\EEq
and use 
\BEq
\label{eq:STOrelation}
\gamma m v = \sqrt{\frac{{\cal E}^2}{c^2}-m^2c^2},
\EEq 
then by setting 
\BEq
\rho_{\rm min} = R
\EEq
we can determine the upper energy, ${\cal E}_{\rm max}$, relative to
the neutron star
at which the particle can still be deflected by the star's magnetic field.
Substitution of Eqs.\ (\ref{eq:STOrelation}) and (\ref{eq:MvsB0}) into
Eq.\ (\ref{eq:rhoMin}) gives
\BEq
{\cal E}_{\rm max}=c\sqrt{q^2B_0^2R^2+m^2c^2},
\EEq 
in agreement with the previous estimate. We still need to find the 
deflection angle, $\Delta \varphi$, however, since it is this angle 
that provides information about the energy transfer in the original 
reference frame of the observer. 

%%%%%%%%% BEGIN FIG. 2
\begin{figure}[!ht]
\includegraphics[angle=0,width=1.00\linewidth]{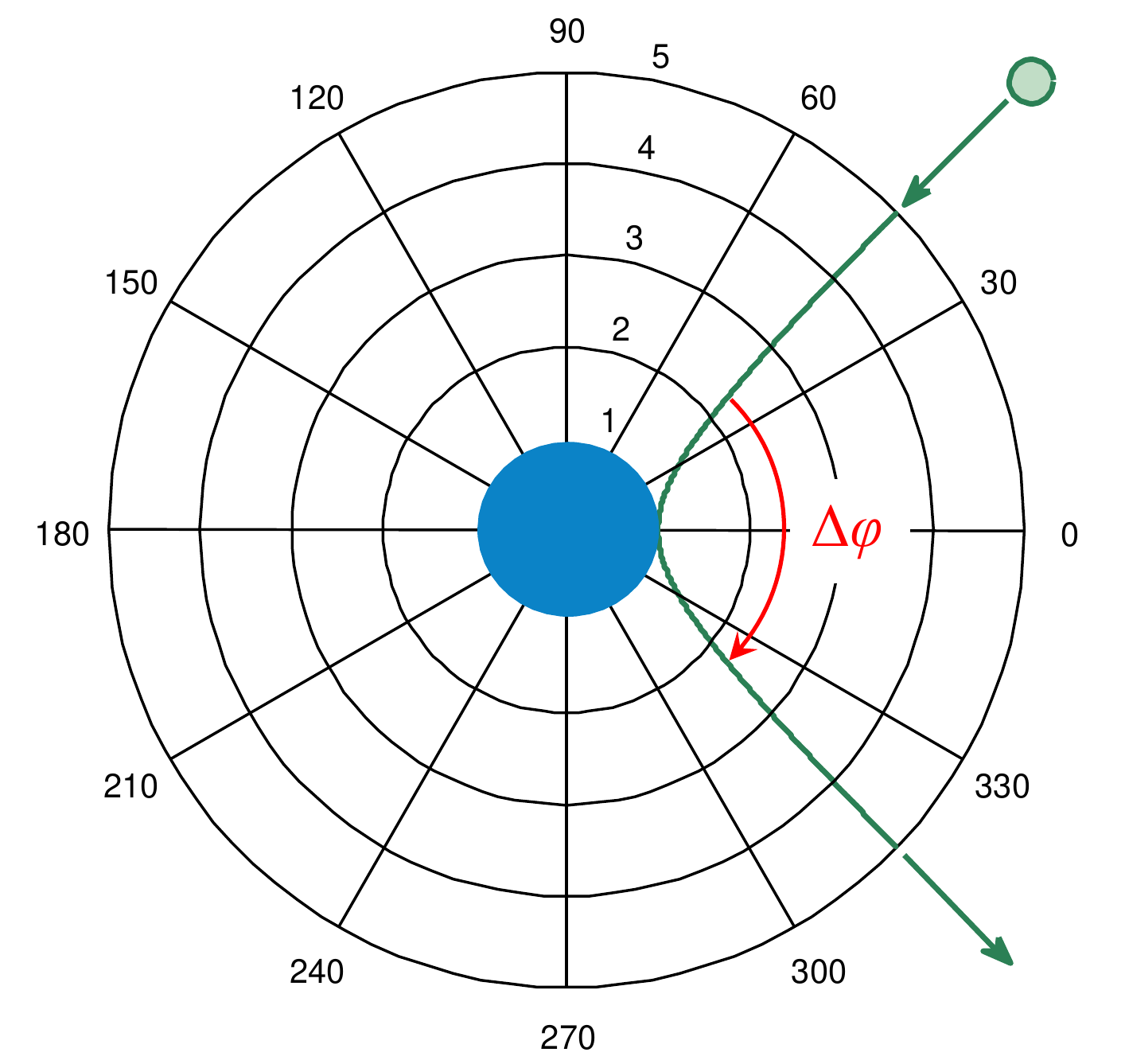}
\caption{ \label{fig:2} 
(color online).
Trajectory of a charged particle (shown in green) 
with energy ${\cal E}_{\rm max}$ 
during a head-on collision with a neutron star (shown in blue), 
as seen from the moving frame of the star.
All distances are measured in units of $\rho_{\rm min}$,
the angles are given in degrees.
}
\end{figure}
%%%%%%%%% END FIG. 2

To find $\Delta \varphi$ for a head-on collision,
we write the conservation equations (\ref{eq:v}) and (\ref{eq:pvarphi}) 
in the dimensionless form,
\BEqA
\label{eq:C1}
\left(\frac{dx}{d \tau}\right)^2 + x^2 \left(\frac{d \varphi}{d \tau} \right)^2
&=& 1,
\\
\label{eq:C2}
x^2 \frac{d \varphi}{d \tau} +\frac{1}{x} 
&=& 0,
\EEqA 
with
$x \equiv \rho/\rho_{\rm min}$, $d\tau \equiv (vdt)/\rho_{\rm min}$.
From Eq.\ (\ref{eq:C2}) we get $d\tau = -x^3 d\varphi$, which
upon substitution into Eq.\ (\ref{eq:C1}) leads to the 
trajectory equation,
\BEq
\frac{d\varphi}{dx} = \pm \frac{1}{x\sqrt{x^4-1}},
\EEq
whose solution is
\BEq
\varphi - \varphi_0 =
\pm \frac{1}{2}\arctan\left(\sqrt{x^4 - 1}\right).
\EEq
The corresponding trajectory at ${\cal E}_{\rm max}$ is depicted in Fig.\ \ref{fig:2}.
Notice that the deflection angle in this head-on collision is 
$\Delta \varphi = \pi/2$,
which means that in the original frame of the observer
the angle is
\BEq
\Delta \varphi_{\rm obs}
=  \arctan 
\left(\frac{mc^2}{{\cal E}_{\rm max}}\right) \ll 1,
\EEq
and the particle will be seen as having energy
\BEq
 {\cal E}_{\rm obs} = \frac{{\cal E}_{\rm max}^2}{mc^2} 
\gg {\cal E}_{\rm max} \gg mc^2.
\EEq
The direction in which the particle will be flying away from the 
impact site will depend on the exact kinematics of the collision. 
In the case of a neutron star radially falling into a Schwarzschild black 
hole, the particle initially present near the black hole will be pushed 
towards the singularity. 
In a more interesting case corresponding to the neutron star slowly spiraling 
into the black hole, the particle may be ejected along a (near) tangential trajectory.
A variation of the latter case may involve the Ban$\tilde{\rm a}$dos-Silk-West mechanism \cite{BSW2009,Jacobson2010,GP2010,{Zaslavskii2010_11_12}}, 
in which both the neutron star and the charge are falling into 
a Kerr black hole from opposite directions.

The above quantitative analysis thus provides justification for our main assertion: 
that direct collisions between neutron stars and black holes may
lead to the production of HE cosmic rays. 
The attractive features of this mechanism include: 
(i) its conceptual simplicity, (ii) its applicability to black holes of wide range of sizes 
(not just supermassive), (iii) the relatively weak magnetic fields involved, and (iv) the possibility of 
cosmic ray generation in chance encounters between rapidly moving neutron 
stars and molecular clouds. The rarity of the high-energy cosmic ray
events is seen 
as a consequence of the accidental nature of all such collisions.

%\section{Conclusions}
%MORE...
%
%\begin{acknowledgments}
%
%MORE...
%
%\end{acknowledgments}

\end{document}